\let\orilabel\label
\documentclass[aps,pra,preprint,a4paper,showpacs,showkeys,superscriptaddress,nofootinbib]{revtex4-1}
\let\label\orilabel

\usepackage{latexsym}
\usepackage{amsmath,amssymb,mathrsfs}
\usepackage{mathtools}
\usepackage{graphicx}
\usepackage{subfigure}
\usepackage{xcolor}
\usepackage{physics}
\usepackage{cancel}
\usepackage{easyReview}
\usepackage{tikz}
\usepackage{multirow,tabularx}

\newcolumntype{C}[1]{>{\centering\arraybackslash}p{#1}}
\usepackage{hyperref}     
\hypersetup{colorlinks,%
  citecolor=blue,%
  linkcolor=cyan,%
}
\usepackage[titletoc]{appendix}
\usepackage{enumerate}
\graphicspath{{./Figures/}}
\usetikzlibrary{decorations.pathmorphing,decorations.pathreplacing,decorations.shapes}
\begin{document}


\title{Investigation of black hole complementarity in AdS\texorpdfstring{$_2$}{(2)} black holes}

\author{Wontae Kim}%
\email[]{wtkim@sogang.ac.kr}%
\affiliation{Department of Physics, Sogang University, Seoul, 04107,
	Republic of Korea}%
\affiliation{Center for Quantum Spacetime, Sogang University, Seoul 04107, Republic of Korea}%

\author{Mungon Nam}%
\email[]{clrchr0909@sogang.ac.kr}%
\affiliation{Department of Physics, Sogang University, Seoul, 04107,
	Republic of Korea}%
\affiliation{Center for Quantum Spacetime, Sogang University, Seoul 04107, Republic of Korea}%
\date{\today}

\begin{abstract}
Black hole complementarity plays a pivotal role in resolving the information loss paradox by treating Hawking radiation as carriers of information, apart from the complicated mechanisms involved in decoding information from this radiation. The thought experiment proposed by Susskind and Thorlacius, as well as the criteria set forth by Hayden and Preskill, provide deep insights into the intricate relationship of black hole complementarity between fiducial and infalling observers. We execute the Alice-Bob thought experiment in the context of two-dimensional anti-de Sitter black holes. It turns out that information cloning can be avoided in the case of a large black hole.
According to the Hayden-Preskill criteria, if the scale parameter associated with the explicit breaking of the one-dimensional group of reparametrizations is significantly exceed the squared mass of the black hole, then information cloning can be effectively evaded.

\end{abstract}
%



\maketitle


\raggedbottom

\section{Introduction}
\label{sec:introduction}
Hawking showed that semi-classical black holes emit thermal radiation and eventually evaporate~\cite{Hawking:1974rv,Hawking:1975vcx}.
This fact raises the question of whether the quantum information contained in the matter that collapsed to form the black hole is lost forever~\cite{Hawking:1976ra}.
Subsequently, several alternative theories have been proposed to resolve the information loss paradox.
These theories encompass the idea that most of the information is emitted through Hawking radiation~\cite{Page:1979tc,tHooft:1990fkf,Susskind:1993if,Susskind:1993ki,Susskind:1993mu,Page:1993wv}, the notion that black holes do not evaporate completely~\cite{Banks:1992ba,Banks:1992is,Banks:1992mi,Giddings:1992ff}, the hypothesis that vast amounts of information are released during the final stages of evaporation~\cite{Hotta:2015yla}, and the introduction of alternative horizon-free and non-singular structures like fuzzballs~\cite{Mathur:2005zp,Skenderis:2008qn}.
A recent alternative viewpoint is also discussed in Ref.~\cite{Unruh:2017uaw}.

In particular, the concept of black hole complementarity (BHC)~\cite{Susskind:1993if} has been discussed to reconcile the Hawking radiation carrying information with the equivalence principle in general relativity.
It suggests that a freely falling observer referred to as Alice crossing the horizon would detect nothing out of the ordinary, while a fixed observer referred to as Bob would detect a significant feature known as the ``stretched horizon.''
In fact, Susskind and Thorlacius~\cite{Susskind:1993mu} noticed that BHC does not contradict the principle of quantum mechanics.
Consequently, they showed that an observer who falls into a black hole after extracting information from the Hawking radiation would not be able to detect the same information inside the black hole before reaching the spacelike singularity.
They assumed that information absorbed by the black hole can be reconstructed after the Page time~\cite{Page:1993wv}.
Additionally, Hayden and Preskill~\cite{Hayden:2007cs} proposed an alternative viewpoint.
They suggested that if information is thrown into a black hole after the Page time, it can still be recovered after the scrambling time, which is typically less than the Page time.
Importantly, Alice and Bob in the black hole interior are also causally disconnected, and thus,
the duplication of information can be evaded.
In connection with BHC, there have also been some related studies~\cite{Kim:2013fv,Chen:2014bva,Gim:2015zra}.

On the other hand, according to AdS/CFT correspondence~\cite{Maldacena:1997re,Witten:1998qj}, the principle of unitarity within conformal field theory strongly implies that observers located in the bulk region should not detect any violations of unitarity.
In relation to BHC, investigations concerning the three-dimensional Ba\~nados–Teitelboim–Zanelli black hole~\cite{Banados:1992wn} have affirmed the validity of BHC~\cite{Gim:2017nnl}.
In this respect, one may question whether BHC is still valid or not in two-dimensional AdS black holes obtained from the Jackiw-Teitelboim (JT) model~\cite{Jackiw:1984je,Teitelboim:1983ux} and the Almheiri-Polchinski (AP) model~\cite{Almheiri:2014cka}.

In this paper, we study the validity of BHC in two-dimensional AdS black holes
after the Page time by executing the thought experiment presented by Susskind and
Thorlacius, and then investigate it in the regime of the Hayden-Preskill's criteria using the scrambling time.
The purpose of this paper is to find out some conditions to evade the duplication of information from two thought experiments distinguished by
the Page time and the scrambling time.
The organization of the paper is as follows.
In Sec.~\ref{Two-dimensional AdS black holes}, we introduce
a dilaton gravity model and explain the singularity structure of the AdS$_2$ black hole with the dilaton field.
For the well-defined thought experiment, we focus on the black hole with the spacelike singularity.
In Sec.~\ref{The thought experiment based on the Page time}, we investigate BHC for the model
based on the Susskind-Thorlacius thought experiment, and find that BHC is safe for a large black hole.
In Sec.~\ref{The thought experiment based on the scrambling time}, we also examine BHC by employing the scrambling time in
the Hayden-Preskill's criteria. We show that if
the scale parameter in the dilaton field is larger than the
squared mass of the black hole, then the cloning of quantum information can be evaded. We give our conclusions and discuss our results
in Sec.~\ref{sec:conclusion}.
In Appendix \ref{sec:appendix}, we discuss the scale parameter in the JT model with the timelike singularity.

\section{Two-dimensional AdS black holes}
\label{Two-dimensional AdS black holes}
Let us consider a two-dimensional dilaton gravity model described by the action
\begin{align}
	\label{def:action}
	S &= \frac{1}{2\pi}\int\dd[2]x \sqrt{-g} \left[ \Phi^2\left( R + \frac{2}{\ell^2} \right) - \frac{2\phi_0}{\ell^2} \right]\\
	  &= \frac{\phi_0}{2\pi}\int\dd[2]x\sqrt{-g}R + \frac{1}{2\pi}\int \dd[2]x\sqrt{-g}\phi\left( R + \frac{2}{\ell^2} \right),
\end{align}
where $\Phi^2=\phi +\phi_0$ represents the dilaton field. Here,
$\phi_0$ is an arbitrary positive constant and $\ell$ is the radius of AdS spacetime.
The dilaton gravity action \eqref{def:action} becomes combination of the topological term and the JT model~\cite{Jackiw:1984je,Teitelboim:1983ux}.
In particular, for $\phi_0 \sim Q^2$ and $\ell \sim Q$ where $Q$ is a magnetic charge, the model can describe the dimensionally reduced
near extremal Reissner-Nordstr\"om (RN) black hole in the near horizon limit~\cite{Maldacena:1998uz}. For $\phi_0=1$, it becomes the AP model~\cite{Almheiri:2014cka}.

In the conformal gauge of $\dd s^2 = -e^{2\rho(u,v)}\dd u\dd v$, the equations of motion are obtained as follows:
\begin{align}
	4\partial_u\partial_v\rho + \frac{1}{\ell^2}e^{2\rho} &= 0,\label{eq:eom phi}\\
	2\partial_u\partial_v \Phi^2 + \frac{1}{\ell^2}(\Phi^2-\phi_0) e^{2\rho} &= 0,\label{eq:eom uv}\\
	\partial_u(e^{-2\rho}\partial_u\Phi^2) = \partial_v(e^{-2\rho}\partial_v\Phi^2) &= 0. \label{eq:eom uu}
	\end{align}
The Liouville equation~\eqref{eq:eom phi} describes the spacetime with a negative constant curvature, and the length element
can be expressed in terms of Poincar\`e coordinates as
\begin{equation}
	\label{eq:metric sol}
	\dd s^2 = -\frac{4\ell^2}{(x^+ - x^-)^2}\dd x^+\dd x^-,
\end{equation}
where the generalized coordinates $x^+(u)$ and $x^-(v)$ are monotonic functions.
From Eq.~\eqref{eq:eom uv} with the constraint equations~\eqref{eq:eom uu}, the dilaton field can be obtained
as
\begin{equation}
	\label{eq:dilaton sol}
	\Phi^2 = \phi_0+ a\frac{1-\kappa Mx^+x^-}{x^+ - x^-},
\end{equation}
where $M$ is the black hole mass and $\kappa = \frac{\pi}{a}$.
 The scale parameter $a$ responsible for explicitly breaking the one-dimensional reparametrization symmetry must remain positive to avoid a strong coupling singularity reaching the boundary within finite proper time~\cite{Almheiri:2014cka}.
Upon using the coordinate transformations of
$x^+ = \frac{1}{\sqrt{\kappa M}}\tanh(\sqrt{\kappa M}u)$ and $ x^- = \frac{1}{\sqrt{\kappa M}}\tanh(\sqrt{\kappa M}v)$,
one can rewrite the solutions~\eqref{eq:metric sol} and \eqref{eq:dilaton sol} in the static form as
\begin{align}
	ds^2 &= -\frac{4\kappa M\ell^2}{\sinh^2(\sqrt{\kappa M}(u-v))}\dd u\dd v,\label{static coordinate}\\
	\Phi^2 &= \phi_0+ a\sqrt{\kappa M}\coth(\sqrt{\kappa M}(u-v)).\label{eq:static dilaton}
\end{align}
Note that in Fig.~\ref{fig:JT}, the metric in Poincaré coordinates~\eqref{eq:metric sol} characterizes the entire black hole spacetime, whereas the metric in static coordinates~\eqref{static coordinate} is confined to describing the exterior region of the black hole.

\begin{center}
	\begin{figure}
		\begin{tikzpicture}[scale=1.2]
			\draw[fill=gray!20!white] (6,6) -- (0,0) -- (6,0);
			\draw[fill=gray!70!white] (6,4) -- (2,0) -- (6,0);
			\draw[->] (6,0) -- (-.25,6+.25) node[above] {\tiny$x^+$};
			\draw[->] (6,0) -- (6+.25,.25) node[above] {\tiny $x^-$};
			\draw[thick] (0,0) -- (0,6);
			\draw[thick] (6,0) -- (6,6);
			\draw[very thick,->] ({4+sqrt(2)/8},{2-sqrt(2)/8}) -- (3,3) node[midway,sloped,below] {\scriptsize Alice};
			\def\x{5.3}\draw[very thick,->] ({5.75+sqrt(2)/8},{3.75-sqrt(2)/8}) -- (\x,9.5-\x) node[midway,sloped,below] {\scriptsize Bob};
			\def\y{.3}\draw[dotted,->,thick] (3.95-\y,2.35+\y) -- (5.3-\y,3.7+\y) node[midway,below,rotate=45] {\scriptsize message};
			\draw[decorate,decoration={snake},->] (4.3,2) -- (5.65,3.35) node[midway,below,rotate=45] {\scriptsize radiation};
		\end{tikzpicture}
		\caption{A Penrose diagram illustrates the AdS$_2$ black hole. 
			In this diagram, both the light and dark gray regions represent the region described by the Poincaré patch \eqref{eq:metric sol}. The dark gray area, in particular, indicates the exterior region of the black hole, which is described by the metric \eqref{static coordinate}.
			}
		\label{fig:JT}
	\end{figure}
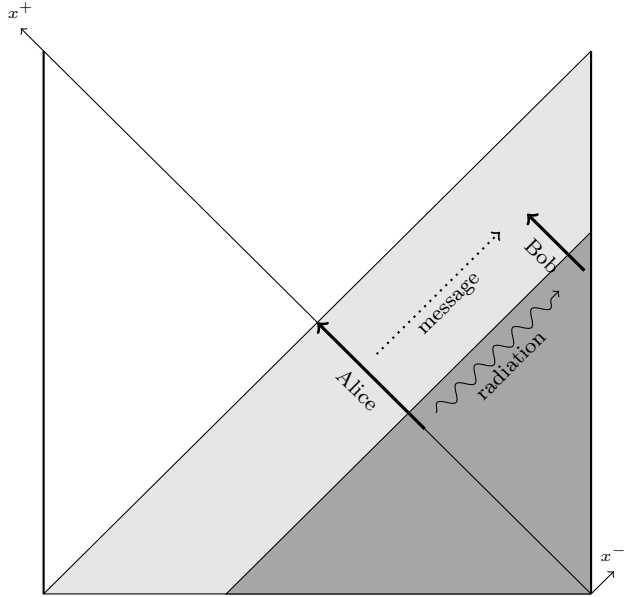
\end{center}

To reveal the singularity structure in the global nature of the geometry,
we consider the global coordinates $y^\pm = y^0 \pm y^1 $. Then, the metric solution takes the form of
\begin{equation}
	\label{eq:global coord}
	\dd s^2 = -\frac{4}{\sin^2 [(y^+ -y^-)/\ell]}\dd y^+ \dd y^-,
\end{equation}
and the dilaton field can be written as
\begin{equation}
	\label{eq:dilaton global}
	\Phi^2 =\phi_0+  a\frac{1-\kappa M\ell^2\tan {(\frac{y^+}{\ell})} \tan {(\frac{y^-}{\ell})}}{\ell[\tan {(\frac{y^+}{\ell})} - \tan {(\frac{y^-}{\ell})}]}.
\end{equation}
The dilaton singularity occurs at
\begin{equation}
	\label{eq:singularity global}
	\tan^2\left(\frac{y^0}{\ell}\right) = \frac{a\kappa M \ell^2\tan^2 (\frac{y^1}{\ell}) + 2\phi_0 \ell\tan{(\frac{y^1}{\ell})} + a}{a\tan^2{(\frac{y^1}{\ell})} - 2\phi_0 \ell\tan{(\frac{y^1}{\ell})}+ a\kappa M\ell^2},
\end{equation}
where $\Phi^2$ becomes zero.
In the Poincar\'e coordinates given by
$x^{\pm} = \ell\tan {(\frac{y^{\pm}}{\ell})}$, the singularity curve~\eqref{eq:singularity global} can be rewritten as
\begin{equation}
	\label{eq:singularity}
	\left( x^+ + \frac{1}{M}\sqrt{\frac{M_c}{\kappa}} \right)\left( x^- - \frac{1}{M}\sqrt{\frac{M_c}{\kappa}}\right) = \frac{M - M_c}{\kappa M^2},
\end{equation}
where $M_c = \frac{\phi_0^2}{\pi a}$.
For $M < M_c$, the singularity becomes timelike and exists near the boundary.
For $ M > M_c$, the singularity becomes spacelike.
Henceforth, we will consider a large black hole where $M \gg  M_c$ to preclude the presence of the timelike singularity.

Note that we will also assume the black hole evolves adiabatically from a thermodynamic perspective. Although the mass reduces to approximately
$(1/4)M$ during evolution, it is still sufficiently larger than $M_c$. Therefore, the spacelike singularity remains as such and does not transform into a timelike singularity within certain time scales, such as the Page time or the scrambling time.

Let us comment on the boundary conditions associated with the AdS black hole.
If the usual reflective boundary condition is chosen for a AdS black hole, the Hawking radiation from the black hole in the bulk can be reflected back into it,
which means that the black hole does not evaporate and remains in thermal equilibrium with its Hawking radiation~\cite{Hawking:1982dh}.
To avoid this, one can choose an absorbing boundary condition by coupling a bulk scalar field representing the Hawking radiation to an auxiliary field at the boundary of AdS~\cite{Rocha:2008fe} since such a coupling permits energy to be transferred from the bulk field to the auxiliary field~\cite{Rocha:2009xy,Rocha:2010zz,Ong:2015fha,Engelsoy:2016xyb,Page:2015rxa}. It is essential to mention the absorption boundary condition for a consistent scenario within the thought experiment. The absorption boundary condition corresponds to the Unruh vacuum condition in the Schwarzschild black hole, which is the best approximation describing an evaporating black hole in a static metric \cite{Unruh:1976db,Unruh:1977ga}: there is no influx at infinity, while the outward flux is also zero at the event horizon. This implies that the outward flux must be perfectly absorbed at the boundary without reflection~\cite{Engelsoy:2016xyb}. Such an account is crucial for rendering the thought experiment a more plausible scenario. In this respect, we will assume the absorption boundary condition at infinity.

\section{The thought experiment based on the Page time}
\label{The thought experiment based on the Page time}
Let us start with the Alice and Bob thought experiment~\cite{Susskind:1993if}. They should be treated as test particles: their energy contributions are assumed to be negligibly small compared to the mass of the black hole, ensuring they do not perturb the metric and dilaton solutions.
Alice falls freely into the horizon of the collapsing black hole, carrying her information.
On the other hand, Bob hovers outside the black hole and collects all the information emitted from the early Hawking radiation.
According to Page's proposal~\cite{Page:1993wv}, Bob can retrieve information about the collapsing matter, including Alice's information, after the Page time has elapsed.
Once this information is extracted from the Hawking radiation, Bob depicted in Fig.~\ref{fig:JT} proceeds to cross the event horizon, moving toward the singularity.
Meanwhile, Alice sends a message encoded with her information to Bob before he reaches the spacelike singularity.
If Bob receives her message before he touches the singularity, he observes the duplication of information and discovers a violation of unitarity in quantum mechanics.
However, this issue may be resolved if the energy of the message Alice sends to Bob requires a scale far beyond the Planckian order.
In other words, the backreaction caused by the super-Planckian energy significantly interrupts the background black hole geometry.
Consequently, we can infer that Alice would be unable to send her message, thereby preventing any duplication of information on Bob's part.

Explicitly, the Page time for an evaporating black hole can be obtained using the Stefan-Boltzmann law:
\begin{equation}
	\label{eq:SB law in JT}
	\dv{M}{t} = -\sigma T^2,
\end{equation}
where $\sigma$ is the Stefan-Boltzmann constant, and the temperature of the black hole is identified with $T = \frac{\sqrt{\kappa M}}{\pi}$.
Note that the thermal entropy is given by
$S_{\rm BH} = \frac{2\pi}{\sqrt{\kappa}}\left( \sqrt{M} + \sqrt{M_c} \right)$~\cite{Almheiri:2014cka},
and this entropy is halved when the mass becomes $M_f = \frac{1}{4}\left( \sqrt{M} - \sqrt{M_c} \right)^2$.
Hence, the Page time can be calculated as
\begin{equation}
	\label{eq:Page time in JT}
	t_{\rm P} = -\frac{\pi^2}{\sigma}\int^{M_f}_{M}\frac{1}{\kappa M}\differential M = -\frac{2\pi^2}{\kappa\sigma} \log\left[ \frac{1}{2}\left(1-\sqrt{\frac{M_c}{M}}\right) \right].
\end{equation}
After the Page time $t_{\rm P}$, Bob enters the black hole so that
\begin{equation}
	\label{eq:Bob x- in JT}
	x^{-}_{\rm B} = \frac{1}{\sqrt{\kappa M}}\tanh(\sqrt{\kappa M}(v_{\rm A} + t_{\rm P})) =  \frac{1}{\sqrt{\kappa M}}\tanh(t_{\rm P}\sqrt{\kappa M}),
\end{equation}
where Alice's initial location $v_{\rm A}$ is set to zero for convenience.
From Eq.~\eqref{eq:singularity}, one can determine the coordinate
when Bob reaches the singularity as $x^{+}_{\rm B} = \frac{a - \phi_0 x^-_{\rm B}}{\pi M x^-_{\rm B} - \phi_0}$.
Hence, if Alice falls freely, the proper time she experiences between the horizon at $x^+ = \frac{1}{\sqrt{\kappa M}}$ and $x^+ = x^+_{\rm B}$ is calculated as
\begin{equation}
	\label{eq:Alice proper time in JT Page}
	\Delta \tau^2 = \ell^2\Delta x_{\rm A}^-\sqrt{\kappa M}\left[ \frac{(\sqrt{ M}+\sqrt{M_c})(1-\tanh{(t_{\rm P}\sqrt{\kappa M})})}{\sqrt{M}\tanh{(t_{\rm P}\sqrt{\kappa M})}-\sqrt{M_c}} \right].
\end{equation}
Using the uncertainty principle $\Delta \tau \Delta E > \frac{1}{2}$, we have
\begin{equation}
	\label{eq:Energy in JT Page}
	\Delta E^2 > \frac{1}{8 \ell^2\Delta x_{\rm A}^-\sqrt{\kappa M}}\left[ \frac{\sqrt{M} - \sqrt{M_c}}{\sqrt{M} + \sqrt{M_c}}e^{2t_{\rm P}\sqrt{\kappa M}} -1 \right],
\end{equation}
where $\Delta x_{\rm A}^- $ is a numerical constant that depends on Alice's initial data \cite{Hayden:2007cs}.
Now, if the mass $M$ is sufficiently greater than $M_c$, the energy uncertainty becomes
\begin{equation}
	\label{important}
	\Delta E^2 > \frac{1}{8 \ell^2\Delta x_{\rm A}^-\sqrt{\kappa M}}e^{2t_{\rm P}\sqrt{\kappa M}} \sim \frac{1}{\sqrt{\kappa M}}e^{2t_{\rm P}\sqrt{\kappa M}} \gg M^2.
\end{equation}
This implies that as long as $M$ is substantially larger than $M_c$, the scale of energy required to convey the message exceeds that of the black hole mass. Therefore, Alice cannot transmit her information to Bob, ensuring that information cloning is avoided.

Note that the ADM mass, denoted as $M$ in Eq.~\eqref{important}, represents the black hole mass as measured by an asymptotic observer. It should correspond to the conserved total energy of the system when considering the negligible energies of Alice and Bob. In the context where Alice remains locally stationary near the horizon, it becomes necessary to account for the gravitational redshift factor. However, since Alice is in a freely falling frame, the introduction of the redshift factor is not required. Thus, employing the conventional uncertainty principle, Alice determines that the energy required for encoding a message into radiation must satisfy $\Delta E \gg M$. It is important to note that the black hole mass on the right-hand side merely implies that the required energy for the message is beyond the total energy of the system. This does not suggest that the mass was measured by the freely falling observer.

\section{The thought experiment based on the scrambling time}
\label{The thought experiment based on the scrambling time}
Let us now assume that the black hole is already maximally entangled with a quantum memory that Bob possesses.
In other words, Bob can decode the initial quantum state encoded within the black hole.
Now, Alice falls freely into the the event horizon of the black hole, her information remaining unknown to Bob.
Meanwhile, Bob remains outside of the black hole and collects all the information emitted from the black hole.
According to the proposal by Hayden and Preskill~\cite{Hayden:2007cs}, Bob can retrieve Alice's information after the scrambling time has elapsed.
As shown in Fig.~\ref{fig:JT}, once Bob retrieves Alice's information from the Hawking radiation, he crosses the event horizon and proceeds toward the singularity.
During this time, Alice attempts to transmit a message encoded with her information to Bob before he reaches the singularity.

Let us execute the thought experiment using the scrambling time.
The scrambling time is given by~\cite{Hayden:2007cs}
\begin{equation}
	\label{eq:Scramb time in JT}
	t_{\rm scr} = \frac{\beta}{2\pi}\log S_{\rm BH} = \frac{1}{2\sqrt{\kappa M}}\log\left[ \frac{2\pi}{\sqrt{\kappa}}\left( \sqrt{M} + \sqrt{M_c} \right) \right],
\end{equation}
where $\beta$ is the inverse of the Hawking temperature.
If Bob crosses the horizon at $x^-_{\rm B}$, then he will reach the singularity at
$x^{+}_{\rm B} = \frac{a - \phi_0 x^-_{\rm B}}{\pi M x^-_{\rm B} - \phi_0}$.
Since Bob jumps into the horizon after the scrambling time $t_{\rm scr}$, his coordinate must satisfy
\begin{equation}
	\label{eq:Bob x- in JT spacetime}
	x^{-}_{\rm B} = \frac{1}{\sqrt{\kappa M}}\tanh(\sqrt{\kappa M}(v_{\rm A} + t_{\rm scr})) =  \frac{1}{\sqrt{\kappa M}}\tanh(\sqrt{\kappa M}t_{\rm scr}),
\end{equation}
where $v_A$ is set to zero just as in the previous section.
The proper time $\Delta \tau$ from the horizon
at $x^{+} = x^+_{\rm A}$ to $x^+ = x^+_{\rm B}$ can be calculated from the metric \eqref{eq:metric sol} near the horizon as
\begin{equation}
	\label{eq:Alice proper time in JT scramb}
	\Delta \tau^2 = \ell^2\Delta x_{\rm A}^-\sqrt{\kappa M}\left[ \frac{(\sqrt{ M}+\sqrt{M_c})(1-\tanh{(t_{\rm scr}\sqrt{\kappa M})})}{\sqrt{M}\tanh{(t_{\rm scr}\sqrt{\kappa M})}-\sqrt{M_c}} \right].
\end{equation}
Using the uncertainty principle, we obtain
\begin{equation}
	\label{eq:Energy in JT scramb}
	\Delta E^2 > \frac{a}{4 \ell^2\Delta x_{\rm A}^-}\left( 1- \sqrt{\frac{M_c}{M}} -\frac{1}{2\pi}\sqrt{\frac{\kappa}{M}} \right).
\end{equation}
To preclude the duplication of quantum information, we particularly assume that
\begin{equation}
	\label{eq:inequality for a}
	a \gg   4 M^2 \ell^2 \Delta x_{\rm A}^- \sim M^2
\end{equation}
in the large black hole. 
Then from Eq.~\eqref{eq:Energy in JT scramb}, we can obtain
\begin{equation}
\Delta E^2 \gg  M^2.
\end{equation}
It means that the scale of energy required to convey the message exceeds 
the black hole mass which amounts the total energy of the system as the result in Sec. \ref{The thought experiment based on the Page time}. 
Thus, Alice cannot transmit her information to Bob, ensuring that information cloning is evaded.
Consequently, the Alice-Bob experiment with the Hayden-Preskill criteria also
demonstrates that the duplication of information can be evaded if the scale parameter $a$ is sufficiently larger compared to $M^2$.

To elucidate the role of the constant $a$ in Eq.~\eqref{eq:inequality for a} from a two-dimensional perspective, we consider the partition function in AdS space. It is noteworthy that two-dimensional dilaton gravity possesses measure-zero degrees of freedom in the bulk: $-1$ from the gravity sector and $+1$ from the dilaton sector. Intriguingly, in contrast to the metric field \eqref{eq:metric sol}, the dilaton field \eqref{eq:dilaton sol} plays a pivotal role in the dynamics of the matter contents. Introducing a boundary at infinity allows us to obtain a non-trivial partition function at the leading order of the dilaton field, expressed as $\Phi^2 = \frac{a}{2\epsilon}$~\cite{Maldacena:2016upp},
where $\epsilon$ is a cutoff that regulates divergence at the boundary. The partition function is then given by $Z = \int \mathcal{D}g_{\mu \nu}\mathcal{D}\Phi^2 e^{iS[g_{\mu\nu},\Phi^2]} \approx \int \mathcal{D}g e^{iS_{\rm bdy}[\tau]}$.
Here, the boundary action can be written in terms of the Schwarzian derivative as $S_{\rm bdy} = -\frac{a}{2\pi}\int\dd t\{\tau(t),t\}$
where $\tau(t) = x^0(t)$.
This implies that non-trivial degrees of freedom can reside on the boundary despite the absence of bulk degrees of freedom. The prefactor $a$ in the Schwarzian action represents the scale of explicit symmetry breaking of the one-dimensional group of reparametrizations at the boundary. This interpretation of $a$ is pertinent when regularizing the dilaton asymptotics. Due to the non-vanishing value of $a$ which breaks the local symmetry, the AP model incorporating a boundary exhibits non-trivial degrees of freedom.

\section{Conclusion and discussion}
\label{sec:conclusion}
We have investigated BHC of the two-dimensional AdS black hole in the dilaton gravity model,
executing the thought experiments based on two different times: the Page time and the scrambling time.
For the black hole possessing a sufficiently large mass, duplication of quantum states is prevented after the Page time, as the energy required for transmitting the message would exceed the mass of the black hole.
At the same time, in the Hayden-Preskill criteria,
the scale parameter $a$ is assumed to be sufficiently larger than the squared mass of the black hole
in order to evade the cloning of quantum information. Consequently, we
conclude that the quantum cloning in the two-dimensional AdS black hole
can be evaded for the large black hole with the large scale parameter $a$.

 We have employed the static solutions for our thought experiment. Actually, it would be necessary to consider the quantum back-reaction on the classical solutions \eqref{eq:metric sol} and \eqref{eq:dilaton sol}, particularly the dilaton field \eqref{eq:dilaton sol} in the AP model~\cite{Almheiri:2014cka,Engelsoy:2016xyb}.
Owing to the Hawking radiation as a form of quantum matter, the solutions must be time-dependent. Nonetheless, the black hole mass is substantially large and it decays slowly in semiclassical approximations. Moreover, in the thought experiment for BHC, we need not consider the endpoint of the evaporation of the black hole, where quantum back-reaction becomes significant. Hence, the black hole can be reasonably assumed to be in quasi-equilibrium with the Hawking radiation up to the Page time or the scrambling time. It implies that the black hole evolves adiabatically for the period of interest. In this context, we could study BHC using static solutions, with the mass parameter varying slowly, without resorting to explicit time-dependent solutions. This approach was previously utilized in deriving the Page time~\cite{Page:1993wv}, where the static metric was used to calculate information for the evaporation of the CGHS black hole. In the seminal work on the thought experiment for BHC~\cite{Susskind:1993if}, the black hole mass reduces to one-quarter of its initial mass at the Page time, when the black hole entropy is half of the initial value. Yet, the mass scale remains on the order of $O(M)$, allowing the metric to be treated as static with the reduced mass. Hence, the assumption of adiabatic behavior in large black holes can also be applicable to our thought experiment.

It is worth nothing that in the case of the four-dimensional Schwarzschild black hole,
the only condition required to avoid the duplication of information in the Alice-Bob experiment is to take the large mass limit.
This holds true irrespective of the values chosen for the Page time and the scrambling time.
Alice needs to encode her message into the radiation with the order of the energy $\frac{1}{\sqrt{M}}e^{\frac{M^2}{2}}$ where $M$ is the mass of the Schwarzschild black hole~\cite{Susskind:1993mu}.
Similarly, if Bob jumps into the black hole after the scrambling time has elapsed, Alice needs to encode her message into the radiation with
the order of the energy $\sqrt{M}$~\cite{Hayden:2007cs}. However, in our current model, the large mass limit is not enough to guarantee the no-cloning theorem.
The additional condition is that the scale parameter $a$ must be significantly larger than the squared mass of the black hole.
One might naturally inquire about what is the corresponding parameter to $a$ in the RN side since
the starting action~\eqref{def:action} can be obtained
from the near-horizon limit of the near-extremal RN black hole which reduces to AdS$_2\cross S^2$ \cite{Maldacena:1998uz}.
The constants $M$, $\phi_0$, and $\ell$ in the model
have been identified by the mass $M_{(4)}$, the magnetic charge $Q$, and the Plank length $\ell_P$ in the RN black hole:
$M = M_{(4)} - \frac{Q}{\ell_{\rm P}}$, $\phi_0 = \frac{\pi Q^2}{2}$, and $\ell = Q\ell_{\rm P}$~\cite{Maldacena:1998uz}.
In Appendix~\ref{sec:appendix}, the additional identification of the scale parameter is elaborated as $a = 2\pi Q^3\ell_{\rm P}$.
All these identifications are valid only under the near-extremal condition, which states $M \ll \frac{Q}{\ell_{\rm P}}$, or equivalently $M \ll M_c$, under which a timelike singularity occurs.
In the present thought experiments, however, we have assumed the condition of $M > M_c$, under which a spacelike singularity occurs. Therefore, our conclusions on the prevention of information cloning
in the two-dimensional AdS black hole are pertinent only when $M >M_c$ and do not extend to the geometry of the RN black hole.

\acknowledgments
We thank Sang-Heon Yi for exciting discussions.
This work was supported by the National Research Foundation of Korea(NRF) grant funded by the Korea government(MSIT) (No. NRF-2022R1A2C1002894) and
by Basic Science Research Program through the National Research Foundation of Korea(NRF) funded by the Ministry of Education through the Center for Quantum Spacetime (CQUeST) of Sogang University (NRF-2020R1A6A1A03047877).

\appendix
\section{The scale parameter \texorpdfstring{$a$}{a} from the RN black hole for \texorpdfstring{$M \ll  M_c$}{M<Mc}}
\label{sec:appendix}
We begin with the four-dimensional Einstein-Hilbert-Maxwell action given by
\begin{equation}
	\label{eq:EHM action}
	S_{(4)} = \frac{1}{16\pi \ell_{\rm P}^2}\int\dd[4]{x}\sqrt{-g_{(4)}}(R_{(4)} - \ell_{P}^2 F_{\mu\nu}F^{\mu\nu}),
\end{equation}
where $\ell_{\rm P}$ is the Planck length, $R_{(4)}$ is the four-dimensional scalar curvature, and $F_{\mu\nu}$ is the field strength tensor for the magnetic field, given by $F = Q\sin \theta\ \dd \varphi \wedge \dd \theta$.
Next, we make an ansatz for a spherically symmetric metric as follows:
\begin{equation}
	\label{eq:met ansatz}
	\dd s^2_{(4)} = g_{ab}\dd x^a\dd x^b + \frac{2\ell_{\rm P}^2}{\pi} e^{-2\psi(x^0,x^1)}\dd \Omega^2,
\end{equation}
where $g_{ab}$ is a (1+1)-dimensional metric and $\psi$ is a dilaton field related to the area of the two-sphere as $r^2=\frac{2\ell_{\rm P}^2}{\pi}e^{-2\psi}$.
Substituting the metric ansatz into the action \eqref{eq:EHM action} yields:
\begin{align}
	\label{eq:reduced EHM action}
	S_{(4)} = \frac{1}{2\pi}\int\dd[2]{x} \sqrt{-g}e^{-2\psi}\left[ R + 2(\nabla\psi)^2 + \frac{\pi}{\ell_{\rm P}^2}e^{2\psi} - \frac{\pi^2}{2\ell_{\rm P}^2}Q^2e^{4\psi} \right].
\end{align}
From the equations of motion,
one can get pure AdS$_2$ geometry of $ R = -\frac{2}{Q^2 \ell_{\rm P}^2}$ with the AdS radius $\ell = Q\ell_{\rm P}$, where the dilaton remains constant as $e^{-2\psi_0} = \frac{\pi Q^2}{2}$.
The fluctuated action around the pure AdS is obtained by expanding $ \psi = \psi_0 + \delta \psi $ as follows~\cite{Almheiri:2014cka,Maldacena:2016upp}:
\begin{align}
	\label{eq:dim red JT}
	S_{(4)}= \frac{\phi_0}{2\pi}\int \dd[2]{x}\sqrt{-g}R + \frac{1}{2\pi}\int\dd[2]{x} \sqrt{-g}\phi\left( R + \frac{2}{\ell^2} \right),
\end{align}
where $\phi_0 = e^{-2\psi_0}$, $\phi = -2e^{-2\psi_0}\delta\psi$, $\phi_0 + \phi = \frac{\pi}{2\ell_{\rm P}^2} r^2$, and
$\phi_0 \gg \phi$ for a small perturbation.

On the other hand, let us now consider the four-dimensional RN black hole solution
\begin{align}
	\dd s^2_{(4)} = -\frac{(r-r_+)(r-r_-)}{r^2}\dd t^2 + \frac{r^2}{(r-r_+)(r-r_-)}\dd r^2 + r^2\dd \Omega^2,\label{eq:RN BH}
\end{align}
where $r_\pm = Q\ell_{\rm P} + E\ell_{\rm P}^2 \pm\sqrt{E^2\ell_{\rm P}^4 + 2QE\ell_{\rm P}^3}$.
Here, $E = M_{(4)} - \frac{Q}{\ell_{\rm P}}$ represents the excitation energy above the extremality described by $E = 0$ or $r_+ = r_- = r_0 = Q\ell_{\rm P}$. Then, the metric reduces to
$\dd s^2_{(4)} = -\left( \frac{r-r_0}{r} \right)^2\dd t^2 + \left( \frac{r}{r-r_0} \right)^2\dd r^2 + r^2\dd \Omega^2$.
For a near-extremal case $E \ll \frac{Q}{\ell_{\rm P}}$, the outer horizon becomes $r_+ = Q\ell_{\rm P} + \sqrt{2QE\ell_{\rm P}^3}$.
Next, let us consider the coordinate transformation defined as
\begin{equation}
	\label{eq:schw transf}
	r = r_0 + \sqrt{2QE\ell_{\rm P}^3}\coth(\sqrt{\frac{2E}{Q^3 \ell_{\rm P}}}x)
\end{equation}
and introduce a small cut-off $\epsilon$ of the space to prevent $\phi$ from becoming too large: $\phi(\epsilon)$ is large, but $\phi_0 \gg \phi(\epsilon)$, indicating that we are in the near-extremal region~\cite{Maldacena:2016upp}.
Plugging Eq.~\eqref{eq:schw transf} into the metric \eqref{eq:RN BH}, we get
\begin{equation}
	\label{}
	\dd s^2 = Q^2\ell_{\rm P}^2\left( \frac{2E}{Q^3\ell_{\rm P}}\frac{-\dd t^2 + \dd x^2}{\sinh^2\left( \sqrt{\frac{2E}{Q^3\ell_{\rm P}}}x \right)} + \dd \Omega^2 \right),
\end{equation}
which describes AdS$_2\times S^2$ geometry with the radius $\ell = Q\ell_{\rm P}$.
From Eq.~\eqref{eq:schw transf}, the dilaton field can be deduced as
\begin{equation}
	\label{eq:near ext RN dilaton}
	\Phi^2=e^{-2\psi_0}(1-2\delta \psi) = \frac{\pi Q^2}{2} + \pi Q\sqrt{2QE\ell_{\rm P}}\coth(\sqrt{\frac{2E}{Q^3 \ell_{\rm P}}}x).
\end{equation}
Comparing Eq.~\eqref{eq:near ext RN dilaton} with Eq.~\eqref{eq:static dilaton}, we can finally obtain
\begin{equation}
	\label{eq:static correspondence}
	 a = 2\pi Q^3\ell_{\rm P}
\end{equation}
with the identifications: $M = E = M_{(4)} - \frac{Q}{\ell_{\rm P}}$,  $\phi_0 = \frac{\pi Q^2}{2}$, and $\ell = Q\ell_{\rm P}$. These identifications are valid under the condition $E \ll \frac{Q}{\ell_{\rm P}}$, equivalently $M\ll M_c$, which generates the timelike singularity.


\bibliographystyle{JHEP}       

\bibliography{reference}

\end{document}